\documentclass[aps,prd,superscriptaddress,showpacs,amsmath,amssymb]{aipproc}
\usepackage{graphicx,epsf}
\usepackage{amsfonts}
\usepackage{amssymb}
\usepackage[]{latexsym}
\layoutstyle{6x9}

\newcommand{\be}{\begin{eqnarray}}
\newcommand{\ee}{\end{eqnarray}}

\newcommand{\Eres}{E_{\rm Res}}
\newcommand{\ELres}{(LE)_{\rm Res}}
\newcommand{\beq}{\begin{equation}}
\newcommand{\eeq}{\end{equation}}

\begin{document}
\begin{flushright}
DO-TH-09/12
\end{flushright}
\title{
Explaining LSND using extra-dimensional shortcuts}
\author{Sebastian Hollenberg}{
address={Fakult\"at f\"ur Physik, Technische Universit\"at Dortmund, D-44221 Dortmund, Germany}
,email=
{sebastian.hollenberg@uni-dortmund.de}}
\author{Octavian Micu\footnote{Speaker.}~}{
address={Fakult\"at f\"ur Physik, Technische Universit\"at Dortmund, D-44221 Dortmund, Germany}
,email=
{octavian.micu@tu-dortmund.de}}
\author{Heinrich P\"as}{
address={Fakult\"at f\"ur Physik, Technische Universit\"at Dortmund, D-44221 Dortmund, Germany}
,email=
{heinrich.paes@uni-dortmund.de}}
\author{Thomas J. Weiler}{
address={Department of Physics and Astronomy, Vanderbilt University, Nashville, Tennessee 37235, USA}
,email=
{t.weiler@vanderbilt.edu}}
\begin{abstract}
We explore the possibility to explain the LSND \cite{Aguilar:2001ty} result in the context of extra-dimensional theories. If sterile neutrinos take shortcuts through extra dimensions, this results in altered neutrino dispersion relations. Active-sterile neutrino oscillations thus are modified and new types of resonances occur.
\end{abstract}
%
%
\maketitle



In theories with large extra dimensions, the Standard Model particles are typically confined to the $3+1$-dimensional brane, which is embedded in an extra-dimensional bulk \cite{ArkaniHamed:1998rs, Randall:1999ee}.
Singlets under the gauge group such as gravitons or sterile neutrinos however are allowed to travel freely on the brane as well as in the bulk. While the active neutrinos are confined to the brane, sterile states can take shortcuts through the extra dimension and the active-sterile neutrino oscillations \cite{Pas:2005rb} generate new resonances.
\par\noindent
\newline
There are different ways in which these bulk shortcuts can be realized, one being the case of asymmetrically-warped extra dimensions. Warp factors shrink the space dimensions $x$ parallel to the brane but leave the time and bulk dimension $t$ and $u$ unaffected \cite{Chung:1999xg, Csaki:2000dm}
\begin{eqnarray}
         d\tau^2 = dt^2 - e^{-2ku} dx^2 - du^2 \label{metric}.
     \end{eqnarray}
When the sterile neutrinos take shortcuts through an extra dimension a resonance arises due to an additional phase difference $\delta (Ht)=t \delta H+ H\delta t$.
Thus, in these models there are two sources of phase difference,
the standard one $t\delta H= L\Delta m^2/2E$, and a new one $Ht\,(\delta t/t)$ arising from temporal shortcuts through the bulk available to sterile neutrinos.
The two phase differences may beat against one another to produce resonant oscillations phenomena.
\par
By introducing the shortcut parameter $\epsilon \equiv  (t^{\rm brane}-t^{\rm bulk})/t^{\rm brane}=\delta t/t$ the effective Hamiltonian for the two state system (one active and one sterile neutrino) is given by
\be
        H_{\rm eff}=\frac{\Delta m^2}{4E} \left( \begin{array}{cc} -\cos2\theta & \sin2\theta \\
        \ \ \ \sin2\theta & \cos2\theta \end{array} \right) -E~\frac{\epsilon}{2} \left(\begin{array}{cc} -1 & 0 \\
        0 & 1 \end{array}\right)\,.
    \ee
One notices that the diagonal terms in the effective Hamiltonian cancel out for a resonance energy $\Eres=\sqrt{\frac{\Delta m^2 \cos2\theta}{2\epsilon}}$. The effective mixing angle becomes
\be
        \sin^2 2\tilde\theta = \frac{\sin^2 2\theta}{\sin^22\theta + \cos^2 2\theta
        \left[1-\frac{E^2}{\Eres^2}\right]^2}. \label{tildesin}
    \ee
It can be seen that there are three distinct energy domains. Below the resonance energy one recovers vacuum mixing, at the resonance energy the effective mixing becomes maximal ($\tilde\theta=\pi/4$), while above the resonance energy oscillations are suppressed.
\par
Starting from the asymmetrically-warped metric in Eq.~(\ref{metric}), one needs to first solve the geodesic equations to calculate the time it takes the sterile neutrinos to travel through the bulk. For the sterile neutrinos to leave the brane, they must have a nonzero initial velocity $\dot{u}_0$ along the extra dimension. When this happens, the geodesics are parabolic and they oscillate about the brane. The distance between two consecutive points where a geodesic crosses the brane is proportional to $\dot{u}_0$. Translational invariance is maintained on the brane and the Minkowski metric on the brane assures that Lorentz invariance is maintained on the brane. Therefore, momentum components are conserved on the brane, and we cannot generate a nonzero $\dot{u}_0$ on the brane except as an initial condition. The uncertainty principle applied to the $u$ dimension allows for such a nonzero velocity. Momentum conservation in the $u$-direction is a non-issue, as translational invariance
in the $u$-direction is broken by the brane itself.
For a baseline $L$, as measured on the brane, there are more  geodesic paths which cross the brane at the position of the detector, and the shortcut parameter for each of those paths is given by
\be
\epsilon_{n}(v) &=& 1-\left(\frac{n}{v}\right)\,
       {\rm arcsinh} \left(\frac{v}{n}\right),   \label{epsilonn}
    \ee
where the scaling variable $v\equiv kL/2$ was introduced, and the subscript $n$ refers to neutrinos which enter the detector upon intersecting the brane for the $n^{\rm th}$ time.
\par
These different modes have to be accounted for when calculating the probability of oscillation. Each mode is weighted by the quantum mechanical weight $e^{iS_n}$, with $S_n$ being the classical action for the free particle.
\par
While the initial momenta are mostly on the brane, the uncertainty principle requires a nonzero $p_u$ component as well. Thus we assume a normalized Gau\ss ian distribution for the momentum component along the extra dimension with a width
 $\sigma$ which is related to the thickness of the brane. The distribution can be written in terms of the mode number through its dependence on velocity component $\dot{u}_0$. In order to be able to select only the geodesics which cross the brane at baseline length $L$, the integral over the momenta needs to be approximated with the corresponding sum with a measure $\Delta n$.
\par
When counting for all neutrinos $\Delta n$ equals one, which is an upper bound. When looking only at the neutrinos which cross through the detector, the value of $\Delta n$ is found by varying the action about the classical extremum.
 We account only for deviations  $\Delta S=|S-S_{\rm cl}|$ smaller than or of the order of $\hbar$ from the classical action, because larger variations lead to rapid oscillations and the integration averages to zero.
\par Putting all the pieces together, the probability of oscillation including the weights mentioned above is given by
\be
        P_{\rm as} = \left|\sum_{n=1}^{\infty} \Delta n\;e^{iS_{\rm cl}(n)}\,
        \frac{v n}{(n^2+v^2)^{3/2}}~
        \left[  \sqrt{\frac{2}{\pi}}\,\frac{\beta E}{\sigma}~ e^{-\frac{(\beta E v)^2}{2\sigma^2(n^2+v^2)}}  \right]
        \sin 2\tilde\theta_n \ \sin  \frac{L\delta\tilde{H}_n}{2}\,\right|^2,
        \label{amplitude}
    \ee
with $\sin 2\tilde\theta_n$ and $\delta\tilde H_n$ obtained by replacing $\Eres$ with the resonant energy of the mode $n$.
 This now depends both on the energy $E$ and the baseline $L$. The factor $\beta$ is the velocity of the sterile neutrinos.
\par Particularly interesting is the "Near Zone", defined for values $v/n\ll 1$. A detailed motivation can be found in \cite{Hollenberg:2009ws}. After making the small $v/n$ expansion, a new feature which emerges is that the resonance peaks are functions of the energy and of the baseline through the combination $LE$ rather than the energy alone as in the MSW matter-resonance \cite{Wolfenstein:1977ue, Mikheev:1986wj}. The $LE$ dependence of the resonances is a novel feature of our model.
\par
\begin{figure}
\centering
\raisebox{3.2cm}{${P_{as}}$}
\includegraphics[scale=0.23]{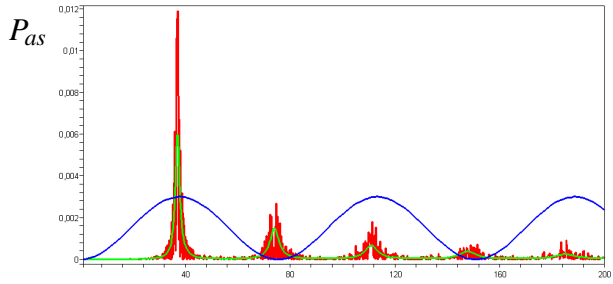}
\caption{Oscillation probability as a function of the experimental baseline, for a Gau{\ss}ian distribution for~$\dot{u}_0$ (red and green curves). The green curve presents the phase-averaged oscillation probability, and the sinusoidal blue curve presents the probability as given by the standard 4D~vacuum formula for oscillations between sterile and active neutrinos. Parameter choices are $\sin^2 2\theta=0.003$, $k=5/(10^{8}~m)$, $E=15$~MeV, $\Delta m^2=64 ~eV^2$,
and $\sigma=100~eV$. The resulting value of $\ELres$ is 550~m~MeV.
For our choice of $E$, the resonance peaks are found at the multiples  $L=n\ELres/E=37n$~m,
$n=1,2,3\cdots$, with the principal resonance corresponding to $n=1$.} \label{fig:gaussian1}
\end{figure}

 The plot in Fig. \ref{fig:gaussian1} shows the probability of oscillation as a function of the baseline for the same value of energy. One can see the consecutive peaks corresponding to consecutive modes $n$. In our case the term $\sin(L\delta\tilde H_n/2)$ oscillates fast, and phase-averaging then sets $\langle \sin\frac{L\delta\tilde H_n}{2}\rangle$ to zero and $\langle \sin^2\frac{L\delta\tilde H_n}{2}\rangle$ to $\frac{1}{2}$.
\par
\begin{figure}
\centering
\raisebox{3cm}{$P_{as}$}
\includegraphics[scale=0.23]{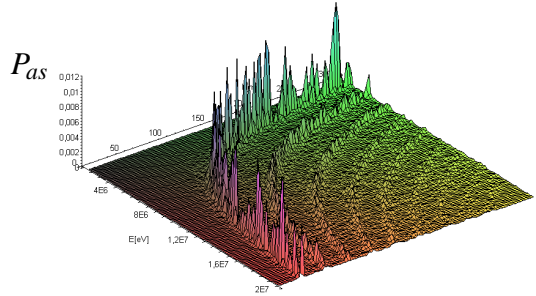}
\caption{Oscillation probability (vertical) in the $L$-$E$~plane, with the same parameters as in Fig.~\ref{fig:gaussian1}.
Units of $L$ and $E$ are m and eV, respectively.} \label{fig:gaussian_distribution3D}
\end{figure}

On the plot in Fig. \ref{fig:gaussian_distribution3D}, the dependence on the product $LE$ of the oscillation probability becomes obvious. The resonance peaks are distributed in hyperbolic patterns on the $L$ vs. $E$ plane, with the peak closest to the origin of the axes being the one for $n=1$. The relative height of consecutive peaks depends on the thickness of the brane $\sigma$. Higher $LE$ resonances are suppressed, and active-sterile neutrino mixing is suppressed for $LE$ above the resonant values.
\par The resonance encountered in our model might explain the observed excess in the LSND data. Sterile neutrinos decouple from the active neutrinos for long-baseline experiments as well as for high energies.
Thus, no active-sterile mixing is expected in atmospheric data, in MINOS \cite{Michael:2006rx} or CDHS \cite{Abramowicz:1984yk}. All explanations proposed  so-far for the LSND and MiniBooNE anomalies assumed baseline-independent oscillations and mixing. Our model relies on metric shortcuts, and it does not discriminate between particles and antiparticles. Thus it will be difficult to accommodate the recent MiniBooNE claims that an excess of flavor changing events exists in the neutrino channel~\cite{AguilarArevalo:2007it} but not in the antineutrino channel~\cite{AguilarArevalo:2009xn}. It might still be possible though to explain the MiniBooNE data by non-standard matter effects \cite{Hollenberg:2009bq}. The failure of previous models to reconcile short baseline data such as LSND with longer baseline data might be construed as favoring the extra-dimensional shortcut scenario. Finally, the bulk shortcut scenario might even relieve some of the remaining tension between the LSND and KARMEN \cite{Armbruster:2002mp} experiments since LSND has almost twice the baseline of KARMEN.



\begin{thebibliography}{99}

\bibitem{Aguilar:2001ty}
  A.~Aguilar {\it et al.}  [LSND Collaboration],
  Phys.\ Rev.\  D {\bf 64}, 112007 (2001)
  [arXiv:hep-ex/0104049].
\bibitem{ArkaniHamed:1998rs}
  N.~Arkani-Hamed, S.~Dimopoulos and G.~R.~Dvali,
  Phys.\ Lett.\  B {\bf 429}, 263 (1998)
  [arXiv:hep-ph/9803315];
  I.~Antoniadis, N.~Arkani-Hamed, S.~Dimopoulos and G.~R.~Dvali,
  Phys.\ Lett.\  B {\bf 436}, 257 (1998)
  [arXiv:hep-ph/9804398];
  N.~Arkani-Hamed, S.~Dimopoulos and G.~R.~Dvali,
  Phys.\ Rev.\  D {\bf 59}, 086004 (1999)
  [arXiv:hep-ph/9807344].

\bibitem{Randall:1999ee}
  L.~Randall and R.~Sundrum,
  Phys.\ Rev.\ Lett.\  {\bf 83}, 3370 (1999)
  [arXiv:hep-ph/9905221];
  L.~Randall and R.~Sundrum,
  Phys.\ Rev.\ Lett.\  {\bf 83} (1999) 4690
  [arXiv:hep-th/9906064].

\bibitem{Pas:2005rb}
  H.~P\"as, S.~Pakvasa and T.~J.~Weiler,
  Phys.\ Rev.\  D {\bf 72}, 095017 (2005)
  [arXiv:hep-ph/0504096].
\bibitem{Chung:1999xg}
  D.~J.~H.~Chung and K.~Freese,
  Phys.\ Rev.\  D {\bf 62}, 063513 (2000)
  [arXiv:hep-ph/9910235];
  D.~J.~H.~Chung and K.~Freese,
  Phys.\ Rev.\  D {\bf 61}, 023511 (2000)
  [arXiv:hep-ph/9906542].

\bibitem{Csaki:2000dm}
   C.~Csaki, J.~Erlich and C.~Grojean,
   Nucl.\ Phys.\  B {\bf 604}, 312 (2001)
   [arXiv:hep-th/0012143].
  R.~R.~Volkas,
  Prog.\ Part.\ Nucl.\ Phys.\  {\bf 48}, 161 (2002)
  [arXiv:hep-ph/0111326].

\bibitem{Hollenberg:2009ws}
  S.~Hollenberg, O.~Micu, H.~P\"as and T.~J.~Weiler,
  arXiv:0906.0150 [hep-ph].
\bibitem{Wolfenstein:1977ue}
  L.~Wolfenstein,
  Phys.\ Rev.\  D {\bf 17}, 2369 (1978).

\bibitem{Mikheev:1986wj}
  S.~P.~Mikheev and A.~Y.~Smirnov,
  Nuovo Cim.\  C {\bf 9}, 17 (1986);
\bibitem{Michael:2006rx}
  D.~G.~Michael {\it et al.}  [MINOS Collaboration],
  Phys.\ Rev.\ Lett.\  {\bf 97}, 191801 (2006)
  [arXiv:hep-ex/0607088].
\bibitem{Abramowicz:1984yk}
  H.~Abramowicz {\it et al.},
  Z.\ Phys.\  C {\bf 25}, 29 (1984).
\bibitem{AguilarArevalo:2007it}
  A.~A.~Aguilar-Arevalo {\it et al.}  [The MiniBooNE Collaboration],
  Phys.\ Rev.\ Lett.\  {\bf 98}, 231801 (2007)
  [arXiv:0704.1500 [hep-ex]];
  A.~A.~Aguilar-Arevalo {\it et al.}  [MiniBooNE Collaboration],
  Phys.\ Rev.\ Lett.\  {\bf 102} (2009) 101802
  [arXiv:0812.2243 [hep-ex]].
\bibitem{AguilarArevalo:2009xn}
  A.~A.~Aguilar-Arevalo {\it et al.},
  arXiv:0904.1958 [hep-ex].
\bibitem{Hollenberg:2009bq}
  S.~Hollenberg and H.~P\"as,
  arXiv:0904.2167 [hep-ph].
\bibitem{Armbruster:2002mp}
  B.~Armbruster {\it et al.}  [KARMEN Collaboration],
  Phys.\ Rev.\  D {\bf 65}, 112001 (2002)
  [arXiv:hep-ex/0203021].
\end{thebibliography}
\end{document}